\documentstyle[preprint,aps]{revtex}

\begin{document}
\title{A Quantum Logic Gate Representation of Quantum Measurement: Reversing and
Unifying the Two Steps of von Neumann's Model}
\author{Giuseppe Castagnoli}
\address{Elsag spa, Via Puccini 2, 16154 Genova Italy}
\date{\today}
\maketitle

\begin{abstract}
In former work, quantum computation has been shown to be a problem solving
process essentially affected by both the reversible dynamics leading to the
state before measurement, and the logical-mathematical constraints
introduced by quantum measurement (in particular, the constraint that there
is only one measurement outcome). This dual influence, originated by
independent initial and final conditions, justifies the quantum computation
speed-up and is not representable inside dynamics, namely as a one-way
propagation.

In this work, we reformulate von Neumann's model of quantum measurement at
the light of above findings. We embed it in a broader representation based
on the quantum logic gate formalism and capable of describing the interplay
between dynamical and non-dynamical constraints. The two steps of the
original model, namely (1) dynamically reaching a complete entanglement
between pointer and quantum object and (2) enforcing the
one-outcome-constraint, are unified and reversed. By representing step (2)
right from the start, the same dynamics of step (1) yields a probability
distribution of mutually exclusive measurement outcomes. This appears to be
a more accurate and complete representation of quantum measurement.

\end{abstract}

\section{Introduction}

\noindent A former research$^{\left[ 1\right] }$ on the character of the
quantum speed-up (the fact that quantum problem solving can be more
efficient than all known classical algorithms) has identified some very
special features concerning the non-dynamical character of quantum
measurement.

In the current work, these features will be used to reformulate von
Neumann's quantum measurement model in what appears to be a more accurate
and complete way.

It should be noted that the approach pursued in $\left[ 1\right] $ is in
contrast with the dominant notion that ``quantum algorithms'' are in fact
algorithms, namely sequential Turing-machine computation and thus dynamical
processes (see also [2-5]). We believe that this very common notion might
obfuscate understanding why there is the speed-up. In the interpretation
given in $\left[ 1\right] $, ``quantum algorithms'' are not algorithms and
are not dynamical processes. In fact, the speed-up is shown to {\em violate}
dynamics, namely the topical assumption that all time evolutions can be
described as one-way propagations\footnote{%
By this we mean that the evolving state at time $t+dt$ is either a
deterministic or a stochastic function of the state at time $t$.}.

In the following Section, we will review the conceptual points of the
approach in question, which is common to the current work (we refer to [1]
for the mathematics involved). We apologize if we will have to re-examine
some very basic notions of computer science. This is needed to introduce, so
to speak, from scratch a computation paradigm that justifies the speed-up.

\subsection{Quantum algebraic computation}

We shall re-examine the notions of {\em definition} and {\em computation} in
the framework of the arithmetical problems we are dealing with. Of course,
these problems implicitly (algebraically) {\em define} their solutions.
Factorization is an example: given the known product $c$ of two unknown
prime numbers $x$ and $y$, the numerical algebraic equation $c=x\cdot y$
implicitly defines the values of $x$ and $y$ which satisfy it. In order to 
{\em compute} these values, the algebraic definition must be substituted by
an {\em equivalent} constructive definition, namely by an algorithm which
prescribes how to compute the solution. The notion of algorithm must thus
include an abstraction of the way things are constructed in reality. The
Turing machine is an example, the Boolean network representation of a
computation process is another.

For historical reasons, the ``reality'' we are dealing with is the classical
non-relativistic one. Here any construction process (and moreover, any
physical change in time) is generally believed to be representable as a
dynamical process.

Consequently, there is a tight parallelism between algorithms and dynamics.
An algorithm originates a propagation of one-way conditional logical
implications (i.e. the logical state at step $k+1$ is a Boolean function of
the state at step $k$), which starts from a completely defined input to end
up in an output. Similarly, dynamics originates a one-way time-propagation
(i.e. the physical state at time $t+dt$ is a function of the state at time $%
t $), which starts from a completely defined initial state to end up in a
final state. The same definition can hold for non-deterministic algorithms
and stochastic dynamics, provided that the above functions are considered to
be stochastic in character.

We should emphasize the fact that these propagations must be one-way -- we
exclude dynamics on closed time-like lines. Consequently, it is never the
case that the Boolean network representation of a computation process has
undetermined inputs, and outputs either constrained to fixed values or
connected to upstream\footnote{%
With respect to the direction of logical implication which, in a computation
process, is the direction of time.} inputs participating in the
determination of those same outputs (i.e. belonging to ``feedback loops'').

Interestingly, Boolean networks, besides representing the process of
computing the solution of a numerical problem, can represent the problem
itself, namely the algebraic, implicit definition of the solution. For
example, the former equation $x\cdot y=c$ can be represented as a network of
Boolean gates with the undetermined inputs $x$ and $y$ and the constrained
output $c$. Evidently, these latter Boolean networks are not subject to the
constraints applying to the Boolean network representation of computation.

This revisitation of the notion of algorithm will serve to show that the
``quantum algorithms'' do not fit it.\ On the contrary, they do operations
that are forbidden to an algorithm, in fact they {\em violate} the notions
of both algorithm and dynamics.

This is particularly clear in the ``quantum algorithms'' yielding an
exponential speed-up, e.g. in Simon's$^{\left[ 6\right] }$ and Shor's$^{%
\left[ 7\right] }$ algorithms. In such algorithms, at some stage of the
computation process quantum measurement selects, out of the parallel outputs
produced that far, all and only those outputs that correspond to the same
eigenvalue of some observable. This is of course an immediate consequence of
the one-outcome constraint. The mathematics and logics inherent in that
selection (see $\left[ 1\right] $ for details) are equivalent to {\em %
creating and solving} {\em an algebraic system of numerical (i.e. Boolean)
equations representing the problem to be solved or the hard part thereof} 
{\em -- i.e. the implicit definition of the solution}.

In fact, an essential point is that the Boolean network representation of
this system contains feedback loops. It is well known in engineering and
computer science that, because of such loops, these networks are
exponentially hard to solve (the equivalent computational networks are
exponentially larger). The point is that quantum measurement, by selecting
only one outcome, solves them at the same speed as there were no loops --
loops are completely transparent to the measurement process.

This is an evident justification of the speed-up and hints at the
non-algorithmic, non-dynamical character of the quantum measurement stage of
quantum computation.

As a matter of fact, $\left[ 1\right] $ shows that the outcome of a
computation yielding the speed-up is dually affected by {\em independent}
initial and final conditions\footnote{%
The one-outcome constraint is {\em independent} of the former evolution
since it holds unaltered for all evolutions.}, thus {\em it is} {\em not}
originated by a one-way propagation, namely by dynamics. It is originated by
a mathematical interplay between dynamics (yielding the state before
measurement) and the non-dynamical constraints introduced by quantum
measurement (in particular, the constraint that there is only one
measurement outcome -- the fact that this is randomly chosen is irrelevant
for what concerns the speed-up).

In this sense, ``quantum algorithms'' are not algorithms at all. They belong
to a new non-algorithmic and non-dynamical computation paradigm, where there
is identity between algebraic, implicit definition of a solution and its
physical determination. It is natural to call this paradigm ``quantum
algebraic computation''.

We have already noted that the speed-up has nothing to do with the fact that
the measurement outcome is randomly chosen. In an evolution affected by both
the initial and final conditions, quantum non-determinism assures that no
information is sent back in time.

Noticeably, the concrete fruit of algebraic computation, namely the
speed-up, {\em violates} {\em dynamics} (i.e. the assumption that all time
evolutions are one-way propagations). One can see a precedent in the fact
that quantum theory violates Bell's inequalities, i.e. locality.

\subsection{von Neumann's model revisited}

We shall now go to re-thinking the representation of quantum measurement at
the light of the above results. As anticipated, we will use as a baseline
von Neumann's model, whose essential features are reviewed in the following.

Let the preparation be the state $\alpha \left| 0\right\rangle _{q}+\beta
\left| 1\right\rangle _{q}$ of a qubit $q$, represented in the basis of an
immediately subsequent measurement. This state will be thought of as the
result of a former measurement in a different basis. By $\left[ q\right] $
we designate the binary number stored in qubit $q$. Measuring $\left[ q%
\right] $ in the above preparation yields the eigenvalue $\gamma =0$ or $%
\gamma =1$ in a mutually exclusive way. Correspondingly, the state of the
``pointer'' $p$ of the measurement apparatus goes to $\left| 0\right\rangle
_{p}$ or $\left| 1\right\rangle _{p}$ and, we assume, the quantum state goes
to $\left| 0\right\rangle _{q}$ or $\left| 1\right\rangle _{q}$.

von Neumann's model consists of two separate steps (i.e. not unified in a
common representation). In the first step, the measurement process is
represented by a unitary evolution yielding the following transition from
the state before to the state after measurement:

\begin{equation}
\left| 0\right\rangle _{p}\left( \alpha \left| 0\right\rangle _{q}+\beta
\left| 1\right\rangle _{q}\right) \stackrel{U}{\rightarrow }\alpha \left|
0\right\rangle _{p}\left| 0\right\rangle _{q}+\beta \left| 1\right\rangle
_{p}\left| 1\right\rangle _{q},
\end{equation}

\noindent where $\left| 0\right\rangle _{p}$ is the pointer initial state
and $U$ is a unitary transformation. The second step amounts to randomly
selecting, as the measurement outcome, either $\left| 0\right\rangle
_{p}\left| 0\right\rangle _{q}$ or $\left| 1\right\rangle _{p}\left|
1\right\rangle _{q}$ in a mutually exclusive way and with probabilities
respectively $\left| \alpha \right| ^{2}$ or $\left| \beta \right| ^{2}$.

At the light of the results of ref. [1], we assume of course the
non-dynamical character of the one-outcome constraint. An immediate
consequence is that (1), being a purely dynamical description, in no way can
represent this constraint. Therefore, given that this constraint yields
relevant information for the purpose of predicting (i.e. a-priori
describing) the quantum measurement process, (1) cannot be considered a
maximal description of this process.

von Neumann's dynamical model will be completed by embedding it in a
``broader'' model capable of representing the interplay between dynamical
and non-dynamical constraints. The broader model is very congenially based
on the quantum logic gate formalism.

\section{Quantum gate representation of measurement}

\noindent In the next Section II.A we will develop the logical-mathematical
representation of a very simple measurement. In Section II.B we will show
that this representation can scale and be applied to more complex
measurement situations.

\subsection{One qubit in a coherent superposition}

Let $\alpha \left| 0\right\rangle _{q}+\beta \left| 1\right\rangle _{q}$
\noindent be the state before measurement, at time $t_{0}$, of a qubit $q$.
We want to a-priori describe the measurement process by taking into account
the one-outcome constraint. We shall list the ``tools of the trade'' first
(see Fig. 1).

\begin{center}
Fig. 1
\end{center}

The measurement process will be represented by means of the unitary
input-output transformation undergone by a four qubits register. This
transformation is modeled as a quantum Boolean gate. While qubits $p$ and $q$
are supposed to be physical objects -- their meaning is the same as in
Section I -- the ancillary qubits $e$ and $a$ are (unconventionally)
abstract objects required for the {\em description} of the physical process.
Qubit $e$ will be used in the description of the one-outcome constraint and
qubit $a$ will serve to ensure gate reversibility, as clarified further
below.

At time $t_{0}$, we want to represent the a-priori available knowledge that
measurement will choose one value of [$q$], either $\gamma =0$ or $\gamma =1$
in a mutually exclusive way and with probability amplitudes respectively $%
\left| \alpha \right| ^{2}$ or $\left| \beta \right| ^{2}$. Correspondingly,
the state of the pointer $p$ and qubit $q$ will either go to $\left|
0\right\rangle _{p}\left| 0\right\rangle _{q}$ or $\left| 1\right\rangle
_{p}\left| 1\right\rangle _{q}$. Which eigenvalue is not known at time $%
t_{0} $, therefore we must represent it as a {\em stochastic} {\em Boolean
variable} $\gamma $ $(\gamma =0,1)$ with probability distribution

\begin{equation}
p(\gamma =0)=\left| \alpha \right| ^{2}\text{, }p(\gamma =1)=\left| \beta
\right| ^{2}.
\end{equation}

We are dealing with a final constraint which, in current assumptions, is not
completely representable inside gate dynamics. This constraint is introduced
in the description of the measurement process by means of the ancillary
qubit $e$. We define the input state of qubit $e$ as follows:

\begin{equation}
\left| \psi \right\rangle _{e}\equiv \stackrel{\_}{\gamma }\left|
0\right\rangle _{e}+\gamma \left| 1\right\rangle _{e},
\end{equation}

\noindent where $\stackrel{\_}{\gamma }$ is the negation of $\gamma $.%
\footnote{%
State (3) does not need to have the form $\stackrel{\_}{\gamma }\left|
0\right\rangle _{e}+e^{i\delta }\gamma \left| 1\right\rangle _{e}$, where $%
e^{i\delta }$ is a phase factor, since in any case (i.e. for $\gamma =0,1$)
there is only one element in this superposition; this can also be seen
formally: $e^{i\delta }$ disappears in the density matrix representation of
this state (we should keep in mind that $\stackrel{\_}{\gamma }\gamma =0$).}
We can say that the state of qubit $e$, at time $t_{0}$ before measurement,
is the best prediction of what the state of qubit $q$ {\em will be} at time $%
t_{1}$ after measurement. Gate dynamics will have the task of transferring
to the measurement outcome the constraint applied to the input state of
qubit $e,$ thus satisfying the prediction. It is worth discussing the
character of $\left| \psi \right\rangle _{e}$:

\begin{itemize}
\item  $\left| \psi \right\rangle _{e}$ is the most general solution of the
(algebraic) projection equation $P\left| \psi \right\rangle =\gamma \left|
\psi \right\rangle ,$ \noindent where $P=\left| 1\right\rangle
_{e}\left\langle 1\right| _{e}$ and $\left| \psi \right\rangle $ is a
normalized {\em ket variable} belonging to$\ {\cal H}_{e}\equiv span\left\{
\left| 0\right\rangle _{e},\left| 1\right\rangle _{e}\right\} $ (it is the
``unknown'' of the algebraic equation -- see [1] for the relationship
between this and quantum algebraic computation);

\item  this ``emulates'' in ${\cal H}_{e}$ the projection equation that
generates in ${\cal H}_{q}\equiv span\left\{ \left| 0\right\rangle
_{q},\left| 1\right\rangle _{q}\right\} $ the eigenvalues/eigenstates of the
measurement basis, namely the possible measurement outcomes; from the
standpoint of determining the eigenvalues of $\gamma $, the result is the
same;

\item  since $\gamma $ is a stochastic variable, (3) is an incompletely
defined state (which eigenvalue will be sorted out is not a-priori known),
represented with the method of random phases$^{\left[ 8\right] }$\footnote{%
This method\ is used in place of the usual density matrix representation.
The density matrix of qubit $e$ is obtained by taking the average over $%
\gamma $ of $\left| \psi \right\rangle _{e}\left\langle \psi \right| _{e}$,
which yields $\left\langle \left| \psi \right\rangle _{e}\left\langle \psi
\right| _{e}\right\rangle _{\gamma }=\left| \alpha \right| ^{2}\left|
0\right\rangle _{e}\left\langle 0\right| _{e}+\left| \beta \right|
^{2}\left| 1\right\rangle _{e}\left\langle 1\right| _{e}$ as expected. We
should note that the mutual exclusivity between eigenstates is lost in the
density matrix representation.};

\item  we should not think that qubit $e$ is physically implementable. This
constrained qubit serves to represent a physical principle (in fact, the
one-outcome constraint) and, like all such principles, is abstract in
character.
\end{itemize}

\bigskip

Now we need to define the gate truth table in such a way that the constraint
applied to the input state of qubit $e$ becomes a constraint to be satisfied
by the measurement outcome. This requires that, if the input state of qubit $%
e$ is $\left| 0\right\rangle _{e}$ ($\left| 1\right\rangle _{e}$), then the
output states of qubits $p$ and $q$ are $\left| 0\right\rangle _{p}$ and $%
\left| 0\right\rangle _{q}$ ($\left| 1\right\rangle _{p}$ and $\left|
1\right\rangle _{q}$). Moreover, it is convenient to assume that the input $%
\left| 0\right\rangle _{e}$ ($\left| 1\right\rangle _{e}$) goes identically
into the output $\left| 0\right\rangle _{e}$ ($\left| 1\right\rangle _{e}$);
thus the meaning of the state of qubit $e$ remains unchanged. This
completely defines the gate truth table but for the output column $a$, which
is defined by the requirement (justified further below) of ensuring gate
reversibility (see table I).

\begin{center}
\begin{tabular}{|l|l|l|l|l|l|l|l|}
\hline
\multicolumn{4}{|l|}{Input} & \multicolumn{4}{|l|}{Output} \\ \hline
$e$ & $p$ & $q$ & $a$ & $e$ & $p$ & $q$ & $a$ \\ \hline
0 & 0 & 0 & 0 & 0 & 0 & 0 & 0 \\ \hline
0 & 0 & 1 & 0 & 0 & 0 & 0 & 1 \\ \hline
1 & 0 & 0 & 0 & 1 & 1 & 1 & 0 \\ \hline
1 & 0 & 1 & 0 & 1 & 1 & 1 & 1 \\ \hline
\end{tabular}

Table I
\end{center}

\noindent We can disregard the remaining 12 rows of the truth table, which
can always be compiled in a way that ensures gate reversibility, thus the
unitarity of the corresponding transformation.

Why we require gate reversibility is justified as follows. Before randomly
choosing the value of $\gamma $, we are dealing with the evolution of a {\em %
closed} system of four qubits. The fact that two qubits are abstract in
character (they encode ``knowledge'' concerning the evolution itself), does
not alter the fact that the evolution of the four qubits cannot loose
information, and should therefore be reversible-unitary in character.

We are now in the position of writing the gate input-output transition:

\noindent 
\begin{equation}
\text{Input}\equiv \left( \stackrel{\_}{\gamma }\left| 0\right\rangle
_{e}+\gamma \left| 1\right\rangle _{e}\right) \left| 0\right\rangle
_{p}\left( \alpha \left| 0\right\rangle _{q}+\beta \left| 1\right\rangle
_{q}\right) \left| 0\right\rangle _{a}\stackrel{U}{\rightarrow }
\end{equation}

\[
\text{Output}\equiv \left( \stackrel{\_}{\gamma }\left| 0\right\rangle
_{e}\left| 0\right\rangle _{p}\left| 0\right\rangle _{q}+\gamma \left|
1\right\rangle _{e}\left| 1\right\rangle _{p}\left| 1\right\rangle
_{q}\right) \left( \alpha \left| 0\right\rangle _{a}+\beta \left|
1\right\rangle _{a}\right) , 
\]

\noindent where $U$ is a unitary transformation whose matrix elements are
immediately derivable from the gate truth table. We can see that the
measurement outcome comprises either $\left| 0\right\rangle _{p}\left|
0\right\rangle _{q}$ or $\left| 1\right\rangle _{p}\left| 1\right\rangle
_{q} $ in a mutually exclusive way with probabilities respectively $\left|
\alpha \right| ^{2}$ and $\left| \beta \right| ^{2}$ (keeping in mind the
definition of $\gamma $).

A difficulty of the original von Neumann's model has disappeared. The usual
entanglement between pointer and quantum system has changed form in a way
which constrains the measurement outcome to be a single one.

The output state of qubit $a$ keeps memory of the preparation of qubit $q$,
as it should be. Otherwise, the information about the phase of this
preparation would be lost in the output, and we are dealing with the
evolution of a closed system which cannot loose information. By means of
qubit $a$, further background information is put in the description of the
measurement process. In fact, measurement theory requires that an ensemble
of identical preparations is compared with the corresponding measurement
outcomes. This implies that the {\em description} of each individual
measurement process includes, after measurement, the memory of the
preparation (for example, this could mean putting the memory of the observer
in the description).

As a final step, we can sort out the value of $\gamma $ according to its
probability distribution. Say it comes out $\gamma =1$. Although this is an
irreversible operation, transition (4) remains reversible (we have to
substitute $\gamma =1$ everywhere):

Input $\equiv \left| 1\right\rangle _{e}\left| 0\right\rangle _{p}\left(
\alpha \left| 0\right\rangle _{q}+\beta \left| 1\right\rangle _{q}\right)
\left| 0\right\rangle _{a}\stackrel{U}{\rightarrow }$

Output $\equiv \left| 1\right\rangle _{e}\left| 1\right\rangle _{p}\left|
1\right\rangle _{q}\left( \alpha \left| 0\right\rangle _{a}+\beta \left|
1\right\rangle _{a}\right) .$

In this idealized representation, irreversibility resides in the random
choice between two reversible histories. By the way, writing the value of $%
\gamma $ at previous times does not mean sending information back in time;
it means making the report of an event after the event has occurred. It is
like saying that, yesterday, the lottery number 123... came out. We are
dealing with a model that faces uncertainty; by fixing the value of $\gamma $%
, the same model changes from being an a-priori description to being a
report.

Let us test the model in the case the preparation is itself an eigenstate of
the measurement basis, say it is $\left| 1\right\rangle _{q}$ (thus, $\alpha
=0$, $\beta =1$). This implies that the probability distribution is $%
p(\gamma =0)=0$, $p(\gamma =1)=1$, which yields $\gamma =1$ (thus $\left|
\psi \right\rangle _{e}=\left| 1\right\rangle _{e}$, from eq. 3). By
substituting these values in (4) (or, alternatively, by using the thruth
table row of input $e=1$ and $q=1$), we obtain:

Input $\equiv \left| 1\right\rangle _{e}\left| 0\right\rangle _{p}\left|
1\right\rangle _{q}\left| 0\right\rangle _{a}\stackrel{U}{\rightarrow }$

Output $\equiv \left| 1\right\rangle _{e}\left| 1\right\rangle _{p}\left|
1\right\rangle _{q}\left| 1\right\rangle _{a},$

\noindent as expected.

We should like to make a few comments on the model developed so far:

\medskip

\noindent i) The method of representing together dynamical and non-dynamical
constraints yields a peculiar form of entanglement. From the one hand, one
can say that the first factor in the measurement outcome (4) is not
entangled since, if $\gamma =1$ (0), then $\stackrel{\_}{\gamma }=0$ (1),
anyhow there is only one element in the superposition. From the other, we do
not know whether $\gamma =0$ or $1$ before having performed measurement. In
the face of this uncertainty, we should consider this factor an entangled
state. Naturally, this kind of entanglement yields the appropriate
correlation between the final state of the quantum system and that of the
classical pointer.

\medskip

\noindent ii) It is interesting to examine in more detail how the
non-dynamical constraint on the ancilla $e$ interplays with von Neumann's
dynamics. For this purpose, it is convenient to introduce a more suitable
von Neumann's model, which is directly comparable to transition (4), as
follows:

\begin{eqnarray}
&&\left( \alpha \left| 0\right\rangle _{e}+\beta \left| 1\right\rangle
_{e}\right) \left| 0\right\rangle _{p}\left( \alpha \left| 0\right\rangle
_{q}+\beta \left| 1\right\rangle _{q}\right) \left| 0\right\rangle _{a}%
\stackrel{U}{\rightarrow } \\
&&\left( \alpha \left| 0\right\rangle _{e}\left| 0\right\rangle _{p}\left|
0\right\rangle _{q}+\beta \left| 1\right\rangle _{e}\left| 1\right\rangle
_{p}\left| 1\right\rangle _{q}\right) \left( \alpha \left| 0\right\rangle
_{a}+\beta \left| 1\right\rangle _{a}\right)  \nonumber
\end{eqnarray}

\noindent where $U$ is the same as in (4). Now we should think that all
qubits are ``real'', whereas qubits $e$ and $q$ have been prepared in the
same state. We shall assume that transition (5) is the dynamical step of von
Neumann's model in the case of an ad-hoc designed measurement.\footnote{%
Keeping in mind truth table I, the prescription would be: measure $\left[ e%
\right] $ in preparation (5), ``write'' the measurement outcome in both the
state of the pointer $p$ and that of qubit $q$, keep memory of the input
state of qubit $q$ in the output state of qubit $a$.} Transition (5) can be
directly related to transition (4). In fact, by multiplying the\ two
end-states of (5) by the operator $S\equiv \frac{\stackrel{\_}{\gamma }}{%
\alpha }\left| 0\right\rangle _{e}\left\langle 0\right| _{e}+\frac{\gamma }{%
\beta }\left| 1\right\rangle _{e}\left\langle 1\right| _{e}$, we obtain the
two end-states of (4). This change does not affect gate dynamics ($S$
commutes with the gate transformation), therefore we can assume that
transition (4) follows the same dynamics of von Neumann's model (5).

We can thus state that reversibly reaching a state of maximal entanglement
between qubits $e$, $p$, $q$ (as in transition 5), in the case that one
qubit ($p$) describes the state of a classical object, actually amounts to
having sorted out one measurement outcome (as in transition 4). This
fundamental feature appears to be {\em represented} in the current model
(see also points iii and iv).

\medskip

\noindent iii) In a problem solving context, ref. [1] shows that the action
of measurement at the same time introduces and solves an algebraic system of
Boolean equations; solving this system in the classical framework could be
computationally hard (see Section I). Therefore, the above statement (point
ii) tells that the measurement process is unaffected by the computational
hardness of the logical operations involved by it. A feature which is
essential in order to achieve the quantum speed-up is thus represented in
the current model. The same feature is of course present in von Neumann's
model, but in that case it is postulated, so to speak, from outside the
model.

\medskip

\noindent iv) Until now quantum measurement has been seen as the discrete
transition from the state before to the state after measurement, while von
Neumann's model is continuous in time. It may be interesting to see the
continuous form of the current model.

We should first find the continuous, reversible von Neumann's transformation
corresponding to transition (5) (let us think that this is routine), then
apply the operator $S$ (point ii) to the evolving ket of this
transformation. The result would be a continuous transformation
corresponding to transition (4). For the sake of exemplification, we shall
figure out a possible form of this latter transformation:

\begin{equation}
\left| \psi ,t\right\rangle _{e,p,q,a}=\bar{\gamma}\left| 0\right\rangle
_{e}\left( \alpha \left| 0\right\rangle _{p}\left| 0\right\rangle _{q}\left|
0\right\rangle _{a}+\beta \cos \omega t\left| 0\right\rangle _{p}\left|
1\right\rangle _{q}\left| 0\right\rangle _{a}+\beta \sin \omega t\left|
0\right\rangle _{p}\left| 0\right\rangle _{q}\left| 1\right\rangle
_{a}\right) +
\end{equation}

\[
\gamma \left| 1\right\rangle _{e}\left( \alpha \cos \omega t\left|
0\right\rangle _{p}\left| 0\right\rangle _{q}\left| 0\right\rangle
_{a}+\beta \cos \omega t\left| 0\right\rangle _{p}\left| 1\right\rangle
_{q}\left| 0\right\rangle _{a}+\alpha \sin \omega t\left| 1\right\rangle
_{p}\left| 1\right\rangle _{q}\left| 0\right\rangle _{a}+\beta \sin \omega
t\left| 1\right\rangle _{p}\left| 1\right\rangle _{q}\left| 1\right\rangle
_{a}\right) , 
\]

\noindent where $t$ ranges from $0$ to $\frac{\pi }{2\omega }$. For $t=0,%
\frac{\pi }{2\omega }$, (6) overlaps on the two end-states of (4). As can be
seen, $\stackrel{\_}{\gamma }\left| 0\right\rangle _{e}$ labels a unitary
transformation of the preparation into one measurement outcome, as it should
be (this is already a property of transition 4, as readily checked);
similarly, $\gamma \left| 1\right\rangle _{e}$ labels a unitary
transformation of the same preparation into the other outcome. The
measurement process is thus represented by a probability distribution of
mutually exclusive unitary transformations affected by both ends, each
leading from the preparation to one of the possible measurement outcomes.

By taking the partial trace over $e,p,a$ of $\left| \psi ,t\right\rangle
_{e,p,q,a}\left\langle \psi ,t\right| _{e,p,q,a}$, we obtain the density
matrix of qubit $q$ as\ a function of $\gamma $ (averaging over $\gamma $
would yield the conventional density matrix):

\[
\rho _{q}(t,\gamma )=\bar{\gamma}\left( 
\begin{array}{cc}
\left| \alpha \right| ^{2}+\left| \beta \right| ^{2}\sin ^{2}\omega t, & 
\alpha \beta ^{\ast }\cos \omega t \\ 
\alpha ^{\ast }\beta \cos \omega t, & \left| \beta \right| ^{2}\cos
^{2}\omega t
\end{array}
\right) +\gamma \left( 
\begin{array}{cc}
\left| \alpha \right| ^{2}\cos ^{2}\omega t, & \alpha \beta ^{\ast }\cos
^{2}\omega t \\ 
\alpha ^{\ast }\beta \cos ^{2}\omega t, & \left| \beta \right| ^{2}\cos
^{2}\omega t+\sin ^{2}\omega t
\end{array}
\right) 
\]

Of particular interest are the following end-values (keeping in mind that $%
\gamma +\bar{\gamma}=1$):

\[
\rho _{q}(0,\gamma )=\left( 
\begin{array}{cc}
\left| \alpha \right| ^{2}, & \alpha \beta ^{\ast } \\ 
\alpha ^{\ast }\beta , & \left| \beta \right| ^{2}
\end{array}
\right) ,\text{ }\rho _{q}(\frac{\pi }{2\omega },\gamma )=\left( 
\begin{array}{cc}
\bar{\gamma} & 0 \\ 
0 & \gamma
\end{array}
\right) . 
\]

\noindent We can see that the eigenstates corresponding to the population
elements of the density matrix become mutually exclusive when the coherence
elements vanish. It is natural to think that this model should also be
applicable to the decoherence representation of quantum measurement.

\subsection{Two qubits in a singlet state}

We shall apply the quantum gate representation of measurement to a more
complex situation. Let us consider two qubits, $q_{1}$ and $q_{2}$, in a
singlet state:

\begin{equation}
\frac{1}{\sqrt{2}}\left( \left| 0\right\rangle _{q_{1}}\left| 1\right\rangle
_{q_{2}}-\left| 1\right\rangle _{q_{1}}\left| 0\right\rangle _{q_{2}}\right)
.
\end{equation}

\noindent Here both qubits are represented in the same measurement
reference. By rotating one reference by $\varphi $, state (7) changes into

\begin{equation}
\left| \psi _{\varphi }\right\rangle _{q_{1,}q_{2}}\equiv \frac{1}{\sqrt{2}}%
\left( -\sin \varphi \left| 0\right\rangle _{q_{1}}\left| 0\right\rangle
_{q_{2}}+\cos \varphi \left| 0\right\rangle _{q_{1}}\left| 1\right\rangle
_{q_{2}}-\cos \varphi \left| 1\right\rangle _{q_{1}}\left| 0\right\rangle
_{q_{2}}-\sin \varphi \left| 1\right\rangle _{q_{1}}\left| 1\right\rangle
_{q_{2}}\right) .
\end{equation}

\noindent State (8) will be the preparation of qubits $q_{1}$and $q_{2}$ at
the input of a quantum gate representing measurement of $\left[ q_{1}\right] 
$ and $\left[ q_{2}\right] $. Now the register is 8 qubits: $%
e_{1},e_{2},p_{1},p_{2},q_{1},q_{2},a_{1},a_{2}$. The gate truth table is
just the combination of two independent truth tables of the form I, labeled
by subfixes respectively 1 and 2. Without going into detail, let us develop
the input-output transformation.

The state of input $e$ (Section II.A) is substituted by the product

\begin{eqnarray}
\left| \psi \right\rangle _{e_{1},e_{2}} &=&\left( \stackrel{\_}{\gamma }%
_{1}\left| 0\right\rangle _{e_{1}}+\gamma _{1}\left| 1\right\rangle
_{e_{1}}\right) \left( \stackrel{\_}{\gamma }_{2}\left| 0\right\rangle
_{e_{2}}+\gamma _{2}\left| 1\right\rangle _{e_{2}}\right) =  \nonumber \\
&&\stackrel{\_}{\gamma }_{1}\stackrel{\_}{\gamma _{2}}\left| 0\right\rangle
_{e_{1}}\left| 0\right\rangle _{e_{2}}+\stackrel{\_}{\gamma }_{1}\gamma
_{2}\left| 0\right\rangle _{e_{1}}\left| 1\right\rangle _{e_{2}}+\gamma _{1}%
\stackrel{\_}{\gamma }_{2}\left| 1\right\rangle _{e_{1}}\left|
0\right\rangle _{e_{2}}+\gamma _{1}\gamma _{2}\left| 1\right\rangle
_{e_{1}}\left| 1\right\rangle _{e_{2}}.
\end{eqnarray}

\noindent By comparing (9) with (8), we see that the joint probability
distribution of the two Boolean variables is

\begin{equation}
p\left( \stackrel{\_}{\gamma }_{1}\stackrel{\_}{\gamma _{2}}=1\right)
=p\left( \gamma _{1}\gamma _{2}=1\right) =\frac{1}{2}\sin ^{2}\varphi ,\text{
\ }p\left( \stackrel{\_}{\gamma }_{1}\gamma _{2}=1\right) =p\left( \gamma
_{1}\stackrel{\_}{\gamma }_{2}=1\right) =\frac{1}{2}\cos ^{2}\varphi .
\end{equation}

\noindent Eq. (9), with probability distribution (10), represents the
principle that measuring $\left[ q_{1}\right] $ and $\left[ q_{2}\right] $
in state (8) sorts out a single valuation of the eigenvalues $\gamma _{1}$
and $\gamma _{2}$ with probability equal to the square module of the
corresponding amplitude.

It should be noted that (10) is a complete description of the probability
distribution of $\gamma _{1}$ and $\gamma _{2}$. From it, we can derive any
auxiliary probability distribution; for example: $P\left( \gamma
_{i}=0\right) =P\left( \gamma _{i}=1\right) =\frac{1}{2}$ $\left( \text{for }%
i=1,2\right) $, $P\left( \gamma _{1}=1/\gamma _{2}=1\right) =\sin
^{2}\varphi $ (the sign $/$ means: conditioned to), etc.. These
distributions are useful in the case that measurement of the two qubits is
performed in two successive steps.

The input-output transformation is

\begin{equation}
\text{Input}\equiv \left| \psi \right\rangle _{e_{1},e_{2}}\left|
0\right\rangle _{p_{1}}\left| 0\right\rangle _{p_{2}}\left| \psi _{\varphi
}\right\rangle _{q_{1},q_{2}}\left| 0\right\rangle _{a_{1}}\left|
0\right\rangle _{a_{2}}\stackrel{U}{\rightarrow }
\end{equation}

\begin{eqnarray*}
\text{Output} &\equiv &(\stackrel{\_}{\gamma _{1}}\stackrel{\_}{\gamma _{2}}%
\left| 0\right\rangle _{e_{1}}\left| 0\right\rangle _{e_{2}}\left|
0\right\rangle _{p_{1}}\left| 0\right\rangle _{p_{2}}\left| 0\right\rangle
_{q_{1}}\left| 0\right\rangle _{q_{2}}+\stackrel{\_}{\gamma }_{1}\gamma
_{2}\left| 0\right\rangle _{e_{1}}\left| 1\right\rangle _{e_{2}}\left|
0\right\rangle _{p_{1}}\left| 1\right\rangle _{p_{2}}\left| 0\right\rangle
_{q_{1}}\left| 1\right\rangle _{q_{2}}+ \\
&&\gamma _{1}\stackrel{\_}{\gamma _{2}}\left| 1\right\rangle _{e_{1}}\left|
0\right\rangle _{e_{2}}\left| 1\right\rangle _{p_{1}}\left| 0\right\rangle
_{p_{2}}\left| 1\right\rangle _{q_{1}}\left| 0\right\rangle _{q_{2}}+\gamma
_{1}\gamma _{2}\left| 1\right\rangle _{e_{1}}\left| 1\right\rangle
_{e_{2}}\left| 1\right\rangle _{p_{1}}\left| 1\right\rangle _{p_{2}}\left|
1\right\rangle _{q_{1}}\left| 1\right\rangle _{q_{2}})\left| \psi _{\varphi
}\right\rangle _{a_{1},a_{2}}
\end{eqnarray*}

\noindent where $\left| \psi _{\varphi }\right\rangle _{a_{1},a_{2}}$ is
obtained from eq. (8) by substituting subfix $q$ with $a$.

We can see from (11) that both in the case of simultaneous and subsequent
measurement of the two qubits, each pointer is in a sharp state and the
probabilities of the different outcomes agree with theory.

As the current model can be applied to quantum measurement in an entangled
state, it can evidently be applied to all ``quantum algorithms''.

\section{Conclusions}

Quantum measurement has been represented as a probability distribution of
mutually exclusive unitary transformations affected by both ends, each
leading from the preparation to one of the possible measurement outcomes.
This two-way influence appears to be ``richer'' than (one-way causality)
dynamics, as it can give the quantum speed-up (ref.[1] and Section I).

Of course the quantum speed-up is not a mathematical curiosity but a
concrete thing, which has the potential of giving huge benefits in practical
computation. Consequently, the fact it implies the violation of dynamics,
appears to be a significant feature.

There is a precedent in the violation of Bell's inequalities, thus of
locality. As well known, this latter violation is experimentally verified.
It can be argued that the experimental verification of a quantum speed-up,
thus of dynamics violation, should be as interesting. This should constitute
a further motivation for implementing a quantum computer.

The fact that the\ transition from the quantum to the classical world has
non-dynamical aspects and, for this reason, can give tangible benefits,
apparently unachievable in the dynamical processes of either world, gives
unforeseen grounds to Bohr's idea that this transition is a central thing in
quantum mechanics. This finding should renovate interest in the quantum
measurement problem, while giving a new perspective for investigating it.

\medskip

Thanks are due to A. Ekert for stimulating discussions and valuable
suggestions.

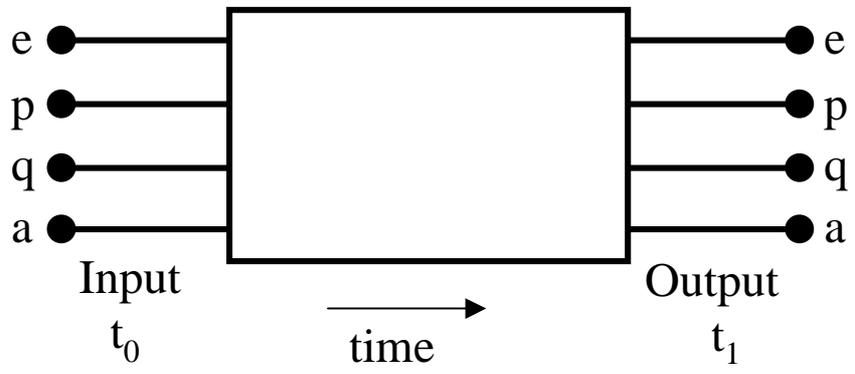
\end{document}